\documentclass[lettersize,journal,colorlinks=true,
linkcolor=black,
urlcolor=blue,
citecolor=blue]{IEEEtran}
\usepackage{amsmath,amsfonts}
\usepackage{algorithmic}
\usepackage{algorithm}
\usepackage{array}
\usepackage[caption=false,font=normalsize,labelfont=sf,textfont=sf]{subfig}
\usepackage{textcomp}
\usepackage{stfloats}
\usepackage{url}
\usepackage{verbatim}
\usepackage{graphicx}
\usepackage{orcidlink}
\usepackage{cite}
\usepackage{siunitx}
\DeclareSIUnit\sq{\ensuremath{\Box}}
\usepackage{cleveref}
\usepackage{layouts}
\usepackage{amssymb}
\usepackage{wasysym} 
\usepackage{soul}
\crefname{figure}{Fig.}{Figs.}
\Crefname{figure}{Fig.}{Figs.}
\crefname{equation}{Eq.}{Eqs.}
\Crefname{equation}{Equation}{Equations}
\usepackage{tikz}
\newcommand\copyrighttext{%
	\footnotesize © 2026 IEEE. Personal use of this material is permitted. Permission from IEEE must be obtained for all other uses, in any current or future media, including reprinting/republishing this material for advertising or promotional purposes, creating new collective works, for resale or redistribution to servers or lists, or reuse of any copyrighted component of this work in other works.}
\newcommand\copyrightnotice{%
	\begin{tikzpicture}[remember picture,overlay]
		\node[anchor=south,yshift=10pt] at (current page.south) {\fbox{\parbox{\dimexpr0.85\textwidth-\fboxsep-\fboxrule\relax}{\copyrighttext}}};
	\end{tikzpicture}%
}

\begin{document}
	
	\bstctlcite{IEEEexample:BSTcontrol} 
	
	\title{Millimeter-Wavelength Lens-Absorber-Coupled Ti/Al Kinetic Inductance Detectors}
	\author{
		Alejandro Pascual Laguna \orcidlink{0000-0001-7293-4150}, Victor Rollano\orcidlink{0000-0001-6878-2609}, Aimar Najarro-Fiandra\orcidlink{0009-0000-9761-2692}, David Rodriguez\orcidlink{0000-0002-0795-7724}, Maria T. Magaz\orcidlink{0000-0002-6008-2971}, \\Daniel Granados\orcidlink{0000-0001-7708-9080}, Alicia Gomez\orcidlink{0000-0002-8752-1401}
		\thanks{Manuscript received 26 September 2025; revised 16 January 2026; accepted 10 February 2026. Date of publication dd Month yyyy; date of current version 16 February 2026. This work was supported by the MCIN/AEI/10.13039/501100011033, the ``NextGenerationEU''/PRTR and the ``ERDF A way of making Europe'' under grants JDC2023-051842-I, PID2022-137779OB-C41, PID2022-137779OB-C42; by the Comunidad de Madrid and the ``NextGenerationEU''/PRTR under grant PR47/21 TAU-CM; and by the Comunidad de Madrid under project Mad4Space-TEC-2024/TEC-182. IMDEA Nanoscience acknowledges financial support from the MCIN/AEI/10.13039/501100011033 under the ``Severo Ochoa'' Program for Centers of Excellence in R\&D (CEX2020-001039-S); and CAB from the Spanish National Research Council (CSIC) Research Platform PTI-001. \emph{(Corresponding authors: Alejandro Pascual Laguna, Alicia Gomez)}}
		\thanks{Alejandro Pascual Laguna is with the Centro de Astrobiología (CSIC-INTA), 28850, Torrejón de Ardoz, Spain, and also with the Delft University of Technology, 2628CD, Delft, The Netherlands. (email: apascual@cab.inta-csic.es)}
		\thanks{Aimar Najarro-Fiandra was with the Universidad de Alcalá, 28801, Alcalá de Henares, Spain, and also with the Centro de Astrobiología (CSIC-INTA), 28850, Torrejón de Ardoz, Spain. He is now with the Delft University of Technology, 2628CD, Delft, The Netherlands. (email: a.najarrofiandra@student.tudelft.nl)}
		\thanks{Victor Rollano, David Rodriguez, Maria T.~Magaz and Alicia Gomez are with the Centro de Astrobiología (CSIC-INTA), 28850, Torrejón de Ardoz, Spain. (email: vrollano@cab.inta-csic.es; drodriguez@cab.inta-csic.es; mmagaz@cab.inta-csic.es; agomez@cab.inta-csic.es)}
		\thanks{D.~Granados is with the Instituto Madrileño de Estudios Avanzados en Nanociencia (IMDEA Nanociencia), 28049, Madrid, Spain. (email: daniel.granados@imdea.org)}
		\thanks{Color versions of one or more figures in this article are available at https://doi.org/10.1109/TASC.2026.3665923.}
		\thanks{Digital Object Identifier 10.1109/TASC.2026.3665923}
	}
	
	\markboth{Transactions in Applied Superconductivity,~vol.~xx, no.~yy, Month~yyyy}%
	{A.~Pascual Laguna \MakeLowercase{\textit{et al.}}: Millimeter-Wavelength Lens-Absorber-Coupled Ti/Al Kinetic Inductance Detectors}


	\maketitle

	\begin{abstract}
		This work presents Ti/Al bi-layer Microwave Kinetic Inductance Detectors (MKIDs) based on lens-coupled spiral absorbers as the quasi-optical coupling mechanism for millimeter-wavelength radiation detection. From simulations, the lens-coupled absorbers provide a 70\% lens aperture efficiency in both polarizations over an octave band with a spiral array absorber and over 10\% relative bandwidth with a single spiral. We have fabricated and measured two devices with bare Ti/Al MKIDs: a $3\times3$ \SI{}{\centi\meter} chip with 9 pixels to characterize the optical response at \SI{85}{\giga\hertz} of the two variations of the absorber; and a large format demonstrator with 253 spiral-array pixels showing potential towards a large format millimeter-wavelength camera. We estimate a sensitivity of \SI{1}{\milli\kelvin\per\sqrt{\hertz}} and a 95\% detector yield.
	\end{abstract}
	
	\begin{IEEEkeywords}
		Absorber, dual-polarization, lens, MKID, millimeter-wavelength, Ti/Al
	\end{IEEEkeywords}
	
	\copyrightnotice
	
	\section{Introduction}
	
	\IEEEPARstart{M}{icrowave} Kinetic Inductance Detectors (MKIDs) \cite{Day2003} are showing very high sensitivities at far-infrared \cite{Baselmans2022,Yates2025}, mid-infrared \cite{Day2024,Dabironezare2025}, and near infrared to optical wavelengths \cite{Szypryt2017,deVisser2021,Kouwenhoven2023}. At millimeter-wavelengths, where coherent detectors have traditionally dominated, instruments based on direct detectors like MKIDs are showing a high technological readiness level for various astronomical applications \cite{NIKA1,NIKA2,KISS,CONCERTO,GroundBIRD,SPT3G,MUSIC,DESHIMA2}. Next-generation millimeter-wavelength instrumentation will rely on large-format focal-plane arrays of MKIDs operating at frequencies down to 30 GHz, with target noise-equivalent powers (NEP) better than \SI{e-17}{\watt\per\sqrt{\hertz}} for future astronomical observatories \cite{FOSSIL,ATLAST}, and ultimate sensitivities approaching \SI{e-20}{\watt\per\sqrt{\hertz}} for sub-\SI{}{\milli\electronvolt} photon-mediated dark matter searches \cite{CADEx}. In order for MKIDs to sense the low energy end of the millimeter wavelengths, low critical temperature ($T_c$) superconductors must be employed. Thin Ti/Al multilayers relying on the proximity effect are most actively being investigated \cite{Catalano2015,Zhao2017,deOry2024}, enabling a tunable gap frequency between 30 and 90 GHz; but other low $T_c$ superconductors like Hf, Ta or Mo could also be considered \cite{Mazin2020}.
	
	Employing low $T_c$ superconductors enables the detection of low energy radiation provided it can reach the detector. To this end, the quasi-optical design must be optimized according to the scientific scenario. Although absorber-coupled detectors generally provide worse angular resolution than antenna-coupled ones \cite{Llombart2018,Kusaka2014}, absorbers are typically easier to fabricate and are more resilient to fabrication tolerances due their insensitivity to phase errors. As opposed to fill-arrays of bare absorbers, lens-coupled absorbers allow a slight decoupling of a proper quasi-optical design with the foreoptics from the detector sensitivity optimization \cite{Dabironezare2025}.
	
	In this article, we propose a broadband lens-absorber-coupled Ti/Al MKID with similar sensitivity in both polarizations, which could be used in millimeter-wavelength polarimetric instruments with the aid of rotating half-wave plates or polarizers \cite{CONCERTO}. We report the cryogenic characterization of the detector optical sensitivity at \SI{85}{\giga\hertz}, and showcase a large millimeter-wavelength MKID camera prototype.
	
	\section{Lens-Coupled Absorber with a Broadband Dual-Polarized Response}
	\label{sec:design}
	
	\IEEEpubidadjcol
	
\begin{figure*}[t]
		\centering
		\includegraphics[]{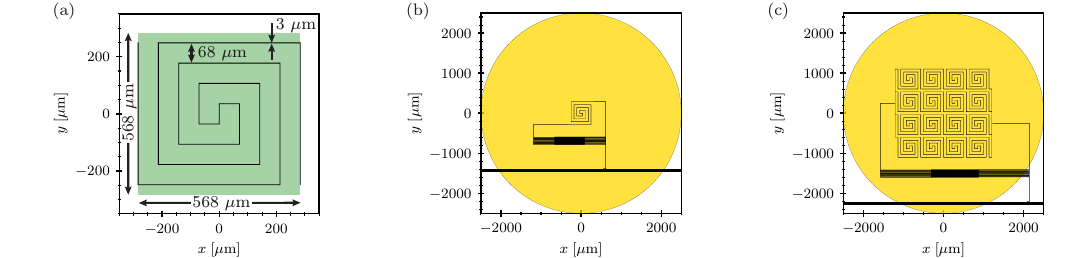}
		\caption{Panel (a) shows the absorbing spiral and its dimensions, where the green square is just a visual aid to show the extent of the unit cell. Panel (b) shows a lumped element resonator with a single spiral as inductive element. Panel (c) shows a resonator with a $4\times4$ spiral array as inductive element. Both designs use an interdigitated capacitor. The yellow circle is the footprint of the lens clear aperture on top of each MKID.}
		\label{fig:resonatorShapes}
	\end{figure*}

	Ultrasensitive broadband dual-polarized absorber-coupled detectors \cite{Echternach2022,Day2024,Dabironezare2025} are key to maximize the observing capabilities of future astronomical observatories. These detectors are typically made of Al because of its ease of fabrication and long quasi-particle lifetimes. However, Al-based absorbers are very difficult to fabricate for sub-millimeter wavelengths, where features become sub-micron to compensate the low impedance of Al. At millimeter wavelengths however, this becomes practical and there is more freedom in the design space. To this end, we propose a Ti/Al ($R_s\approx\SI{1}{\ohm\per\sq} $) absorber consisting of a square double spiral unit cell, with \SI{3}{\micro\meter}-wide arms spiraling inwards and outwards connected in series in the middle. \Cref{fig:resonatorShapes} depicts two lumped-element resonator designs, where the capacitor is an interdigitated arrangement, and the inductor is based on the proposed double spiral unit cell in two variations: (a) with a single spiral and (b) with an array of $4\times4$ spirals whose rows are connected in parallel to avoid self-resonances. The resonators are on a slab of Si backed with a ground plane, serving both as a quarter-wavelength backing reflector for millimeter wavelengths and for the ground plane of the microstrip microwave readout line.
	
\begin{figure*}[t]
		\centering
		\includegraphics[]{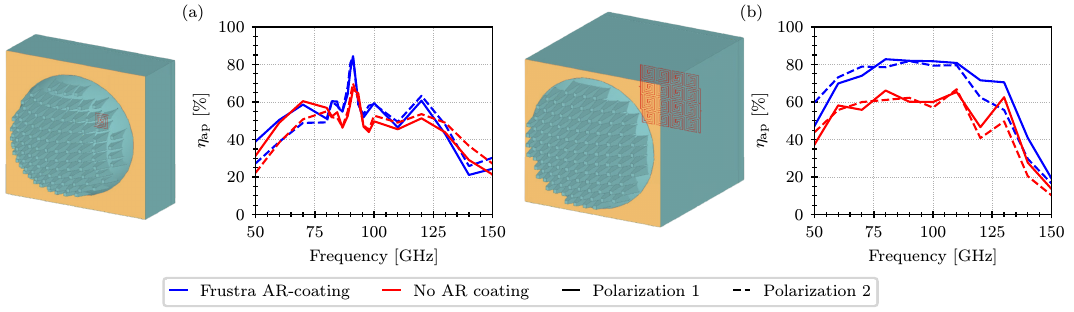}
		\caption{Geometry and simulated aperture efficiency for two orthogonal polarizations for (a) the lens-coupled double-spiral and (b) the lens-coupled double-spiral $4\times4$ array, in both cases analyzed with and without broadband AR coating The absorbers (in red) are at the lower focus of the synthesized elliptical lens. The frustras are not conformal to the lens shape to be able to use current capabilities in laser ablation technology \cite{Bueno2022}. The top surface (in orange) around the lens clear aperture has a sheet impedance of \SI{377}{\ohm\per\sq} to perfectly absorb a free-space plane-wave and it is included in the simulations of the aperture efficiency to limit the input power to the lens clear aperture.}
		\label{fig:spiralArray_drawing_optResp}
	\end{figure*}

	\Cref{fig:spiralArray_drawing_optResp} shows each spiral absorber placed at the lower focus of an extended hemispherical Si lens with a clear aperture diameter of \SI{5}{\milli\meter} and a radius-normalized extension length of $L/R=0.39$ to synthesize an ellipsoidal shape \cite{Filipovic1993}. The single spiral absorber in (a) was coupled to an $f_\#=0.63$ lens to maximize the power collection while avoiding the deepest lens configuration at $f_\#=0.54$, where the lens radius coincides with the clear aperture diameter. This results in a focal field sampling of $1\lambda_\mathrm{Si}f_\#$, where $\lambda_\mathrm{Si}$ is the wavelength in silicon. On the other hand, the $4\times4$ spiral-array in (b) was coupled to an $f_\#=1.2$ lens to maximize the power coupling with the absorber lateral dimension being $2\lambda_\mathrm{Si}f_\#$.
	
\begin{figure}[t]
		\centering
		\includegraphics[]{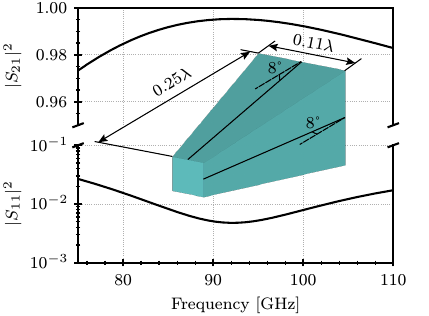}
		\caption{Broadside plane-wave response of the frustra AR-coating unit cell shown in the inset. Port 1 is above the frustra and port 2 is below, embedded in silicon. The dimensions are given as function of $\lambda$, the free-space wavelength at the central frequency of operation, here 92.5 GHz. These dimensions are achievable with current laser ablation technology.}
		\label{fig:AR_coating_wFrustra}
	\end{figure}

	To avoid reflections at the silicon-vacuum interface, the lens top is shaped with vertical (non-conformal) frustras to create a broadband anti-reflection (AR) coating suitable for cryogenic temperatures \cite{Bueno2022}. The design and performance of the AR-coating is reported in \cref{fig:AR_coating_wFrustra} for a broadside plane-wave incidence. The performance of the two lens-coupled absorber designs was simulated with CST Studio Suite \cite{CST} in terms of their broadside plane-wave response for two orthogonal polarizations. In order to restrict the plane wave to the lens clear aperture we added a perfectly absorbing sheet with the free-space impedance. The results of both designs are shown in \cref{fig:spiralArray_drawing_optResp} indicating a lens-aperture efficiency $\eta_\mathrm{ap}$ for both polarizations of more than $70\%$ for about $10\%$ relative bandwidth for the single spiral absorber in (a), and for an octave for the spiral array in (b).
	
	\section{Fabrication}
	\label{sec:fabrication}
	
\begin{figure}
		\centering
		\includegraphics[width=\linewidth]{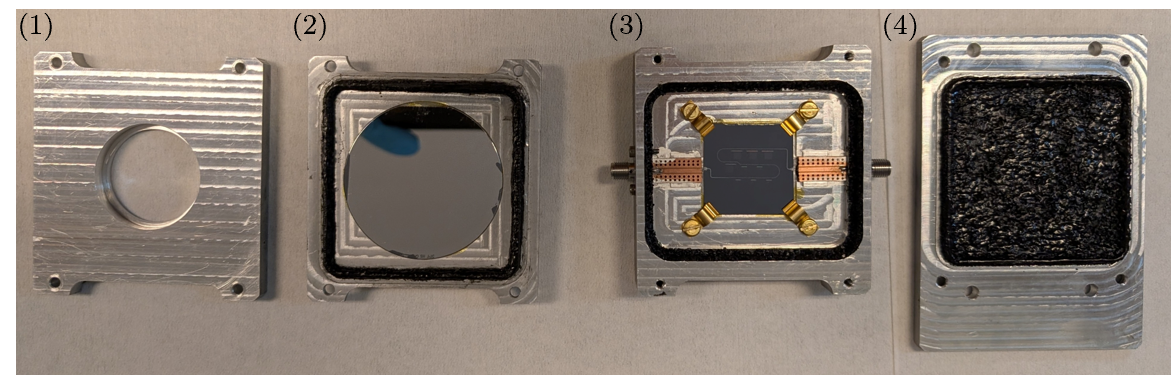}
		\caption{Photograph of the light-tight Al holder for $3\times3$ \SI{}{\centi\meter} chips. The different parts of the holder are: (1) the top lid with an aperture stop of $\diameter$\SI{3}{\centi\meter}, (2) the lid holding a \SI{85}{\giga\hertz} $\diameter$2'' Fabry-Pérot band-pass filter, (3) the part holding the 9 MKID chip wire-bonded to the grounded coplanar waveguide PCB launchers interfacing with the SMA connectors, and (4) the bottom lid to close the assembly. The bottom lid and the joints of the parts are blackened with a thin stray-light absorbing layer.}
		\label{fig:smallDevice}
	\end{figure}

\begin{figure}[]
		\centering
		\includegraphics[width=0.62\linewidth]{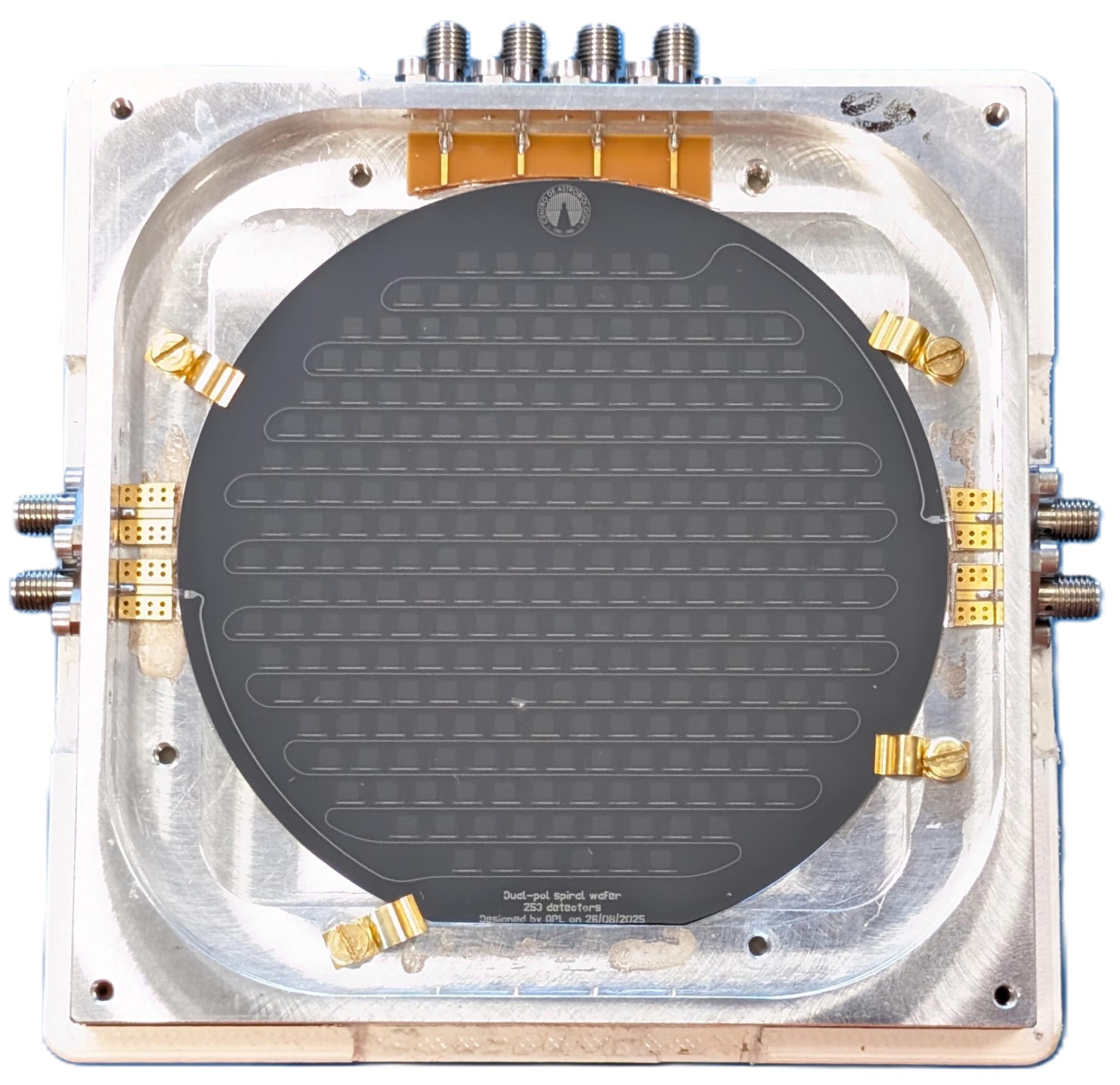}
		\caption{Photograph of the large format demonstrator fabricated on a $\diameter$4'' wafer and hosting 253 MKIDs. The device is placed on an Al holder, where only two of its SMA connectors are employed to read out the device in a frequency-multiplexed fashion.}
		\label{fig:waferDevice}
	\end{figure}

	We have fabricated two lens-less MKID devices: a small $3\times\SI{3}{\centi\meter}$ chip for testing the new designs, comprising 5 spiral array detectors and 4 single spiral detectors; and a $\diameter$4'' large-format demonstrator with 253 spiral array detectors. Both devices were made from high-resistivity ($\rho>\SI{1000}{\micro\ohm\centi\meter}$) double-side-polished float-zone silicon wafers, with a thickness of about \SI{280}{\micro\meter} to realize a quarter-wavelength backing reflector. The recipe is the same for both devices. We first sputter-deposit on the wafer \SI{10}{\nano\meter} of Ti and then, without breaking the vacuum in the sputter chamber to avoid oxidation, we add \SI{20}{\nano\meter} of Al. This layer is patterned to define the resonators and readout line using mask-less laser lithography on a negative resist (AZ nLOF 2070), which is developed in AZ developer. The exposed Ti/Al is then subsequently wet-etched with TechniEtch Al80 to attack Al and TechniEtch TBR19 to attack Ti. To make the ground plane we first protect the frontside with resist and then deposit \SI{200}{\nano\meter} of Al on the backside of the wafer. The fabricated devices can be seen, already assembled in their chip holders, in \cref{fig:smallDevice,fig:waferDevice}. Future experiments will include a Si lens array that is aligned and glued to the detector wafer. The lenses, including the frustra AR-coating, will be manufactured from a monolithic Si wafer by means of laser ablation \cite{Bueno2022}. The glueing of the lens array and device wafers will be done with either a lithographic glue like PermiNex\textsuperscript{\textregistered} or a high viscosity adhesive like Epoxy 2216 \cite{Baselmans2022}.
	
\begin{figure}
		\centering
		\includegraphics[]{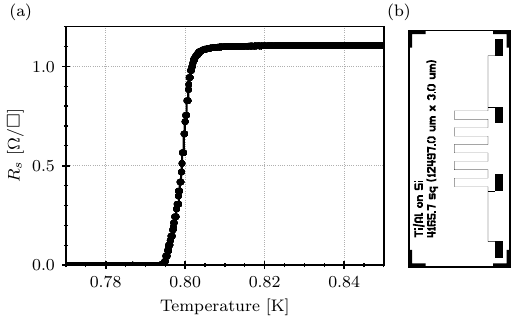}
		\caption{Panel (a) shows the measured sheet resistance of a bi-layer of \SI{11}{\nano\meter}-thick Ti and \SI{19}{\nano\meter}-thick Al. Close to the gap, the normal state sheet resistance is \SI{1.1}{\ohm\per\sq} and the resistivity is \SI{3.3}{\micro\ohm\centi\meter}. The critical temperature is approximately \SI{800}{\milli\kelvin}. Panel (b) shows the layout of the chip employed to measure the resistance.}
		\label{fig:Rs_vs_T}
	\end{figure}

	To assess the Ti/Al bilayer properties, we performed a 4-point DC resistance measurement of a 4165.7 squares-long 15/11 \SI{}{\nano\meter}-thick Ti/Al wire co-fabricated with the small device chip shown in \cref{fig:smallDevice}. The sheet resistance as a function of temperature is given in \cref{fig:Rs_vs_T}, which is \SI{1.1}{\ohm\per\sq} just above the superconducting gap and corresponds to an effective normal state resistivity of $\rho=\SI{3.31}{\micro\ohm\centi\meter}$. The critical temperature is $T_c=\SI{800}{\milli\kelvin}$, resulting in a gap frequency of $f_\mathrm{gap} = 2\Delta_0/h \approx 3.52 k_B T_c / h \approx \SI{59}{\giga\hertz}$, where $k_B$ is Boltzmann's constant and $h$ is Planck's constant. Using the full Mattis-Bardeen equations \cite{MattisBardeen1958} for the complex conductivity and the dirty limit sheet impedance we estimate a kinetic inductance of \SI{1.9}{\pico\henry\per\sq}.
	
\begin{figure}[]
		\centering
		\includegraphics[]{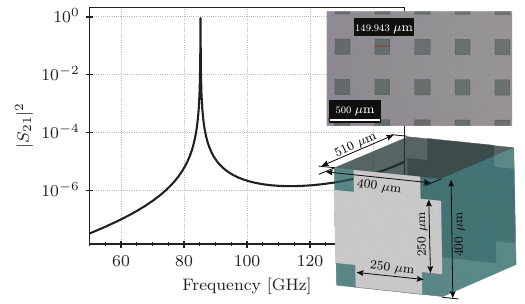}
		\caption{Simulated broadside plane-wave transmission of the perfectly-conducting Fabry-Pérot band-pass filter with the unit cell shown in the lower inset. The layer below is exactly the same as the layer above. The upper inset shows an optical microscope photograph of the fabricated device.}
		\label{fig:FPBPF_withUnitCell}
	\end{figure}

	To perform a controlled optical response characterization of the detectors, we have fabricated the Al chip holder depicted in \cref{fig:smallDevice}, where millimeter-wavelength radiation is only allowed through a $\diameter$\SI{3}{\centi\meter} aperture. All interfaces susceptible of letting stray radiation inside the holder have been carefully engineered to make radiation find a thin absorbing layer composed of a mixture of \SI{1}{\milli\meter} SiC grains, Stycast 2850FT and carbon black powder \cite{Diez2000,Baselmans2022}. Furthermore, in order to ensure the probing of millimeter-wavelength radiation and eventually obtaining the optical efficiency of these multi-moded detectors, we have also designed and fabricated a \SI{85}{\giga\hertz} Fabry-Pérot band-pass filter with the dimensions in \cref{fig:FPBPF_withUnitCell}. It consists of two highly reflective metallic layers separated by a half-wavelength medium. To this end, we use a standard $\diameter$2'' \SI{525}{\micro\meter}-thick high-resistivity ($\rho\gtrsim\SI{1000}{\micro\ohm\centi\meter}$) double-side-polished float-zone silicon wafer. We start by covering the wafer with resist in the backside to protect it for further processing. We subsequently deposit AZ1505 positive resist, and pattern the filter lines with mask-laser lithography. After developing the resist, we evaporate by e-beam  \SI{5}{\nano\meter} of Ti to promote adhesion and, without breaking the vacuum of the evaporator chamber, a \SI{100}{\nano\meter}-thick film of Nb. We finally perform lift-off, thereby removing the metal in the filter gaps. The same steps are repeated on the backside of the wafer with front-to-back alignment using the front and rear cameras of the laser writer. The fabricated filter can be seen mounted in place in the part (2) of the chip holder in \cref{fig:smallDevice}. 
	
	\section{Optical Cryogenic Characterization}
	\label{sec:sensitivity}
	
\begin{figure*}[t]
		\centering
		\includegraphics[]{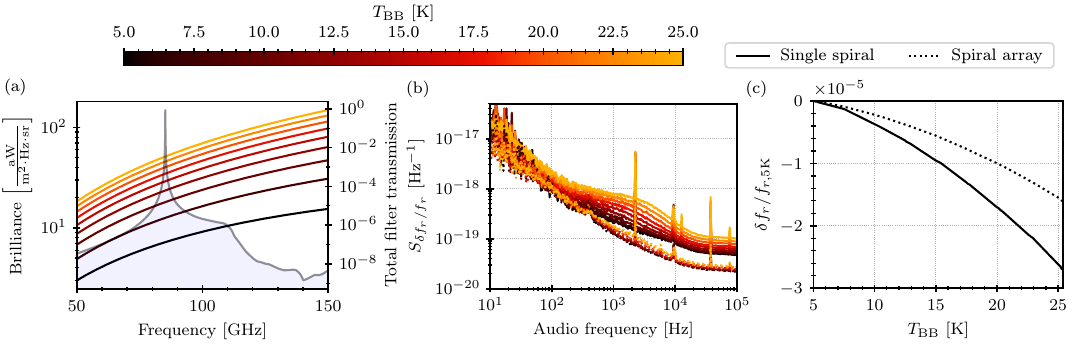}
		\caption{Panel (a) shows the brilliance at the investigated blackbody temperatures and the total filter stack-up transmission. Panel (b) shows the PSD in the relative resonance frequency shift coordinate $S_{\delta{}f_r/f_r}$ for both a single spiral detector and a spiral array detector of the small chip at \SI{100}{\milli\kelvin} bath temperature and with a readout power of \SI{-85}{\deci\bel\milli\relax} at the chip level. Despite the large $1/f$ noise, photon noise is visible for both detectors. Panel (c) shows the response of these two detectors as a function of blackbody temperature.}
		\label{fig:sensitivity}
	\end{figure*}

	The small chip device was cooled to \SI{100}{\milli\kelvin}, and we performed an optical characterization of the two resonator types against a temperature-calibrated blackbody radiator. The blackbody is clamped to the 4K stage of our cryostat and placed directly in front of the chip under test. The blackbody radiation is filtered by a 110 GHz low-pass filter from QMCI \cite{QMCI}, and our own 85 GHz Fabry-Pérot band-pass filter. The blackbody brilliance for the temperatures under consideration $T_\mathrm{BB}$ and the filter stack transmission $F(f)$ are shown in \cref{fig:sensitivity}(a). To obtain the sensitivity of our detectors, we employ the methodology described by Baselmans \emph{et al.} \cite{Baselmans2022}: we stabilize the blackbody at a temperature $T_\mathrm{BB}$, measure the power spectral density (PSD) of the observable variable $x$, sweep the temperature around $T_\mathrm{BB}$ to obtain the responsivity with respect to the blackbody temperature $dx/dP_\mathrm{BB}$. Because we still do not have cryogenic transmission measurements of our filters and the optical throughput is poorly defined in our setup, we only report the response with respect to the blackbody temperature. As can be seen in the PSDs in \cref{fig:sensitivity}(b), our detectors are largely dominated by $1/f$ noise. Despite this excess setup noise, we can clearly see a white spectrum with a roll-off increasing with optical loading, evidence of photon noise \cite{Janssen2013}. The difference in the white noise level is due to the very different quality factor of the particular resonators of choice, but the readout power level at the chip is \SI{-85}{\deci\bel\milli\relax} for both. Furthermore, in \cref{fig:sensitivity}(c) we report the optical response by means of the relative frequency shift $x=\delta{}f_r/f_r$ observable as a function of the blackbody temperature $T_\mathrm{BB}$. Our first estimates for \SI{85}{\giga\hertz} radiation indicate a frequency responsivity of $d(\delta{}f_r/f_r)/dT_\mathrm{BB}\sim\SI{1}{(\hertz\per\hertz)\per\micro\kelvin}$ and thereby a Noise Equivalent Temperature (NET) of order \SI{1}{\milli\kelvin\per\sqrt{\hertz}} for both types of detectors at \SI{1}{\kilo\hertz} audio frequency. The noise equivalent power (NEP) with respect to the impinging power on the detector (i.e.\ including the optical efficiency) is estimated to be of the order of \SI{e-19}{\watt\per\sqrt{\hertz}}, employing the simulated transmission of the quasi-optical filter in \cref{fig:FPBPF_withUnitCell} and assuming an optical throughput of $\lambda^2$; but a more thorough characterization will require a dedicated optimization of the optical setup.
	
	\section{Large Format Demonstrator}
	\label{sec:largeFormatDemonstrator}
	
\begin{figure}
		\centering
		\includegraphics[]{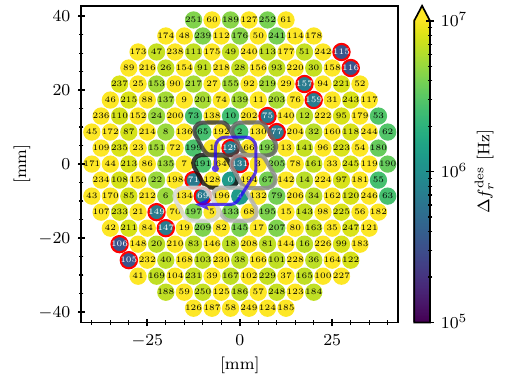}
		\caption{Implemented MKID shuffling strategy, where each circle represents a resonator. The indices indicate the order in frequency space. The grayscale trapezoids exemplify the shuffling strategy, where the 6 first clusters (from darker to brighter) of 4 elements from the 4 different shuffling groups are shown. The color of each circle indicates the closest spectral distance to the directly neighboring resonators, which can be read out from the accompanying colorbar. By design, the average spacing between spatially neighboring resonators is \SI{12.3}{\mega\hertz} and only 14 (5.5\%) resonators are closer than 1 MHz by design to their immediate neighbors (circles with red edges).}
		\label{fig:MKIDshuffling}
	\end{figure}

	Due to the adequate measured optical response of both detector types shown in \cref{sec:sensitivity}, and the very good simulated performance of the $4\times4$ spiral array one described in \cref{sec:design}, we designed and fabricated a large format demonstrator with this detector configuration. One of the biggest challenges to array many detectors in a single readout line is cross-talk between pixels. To mitigate it, the distribution of MKIDs in the wafer follows the shuffling algorithm illustrated in \cref{fig:MKIDshuffling}, which is similar to other encodings in the literature \cite{Noroozian2012,Yates2014}. In this shuffling strategy, the resonance frequencies are split in four contiguous sub-groups across the readout bandwidth and are distributed in cells of four physically neighboring resonators, each of one sub-group. These cells spiral outwards from the array center while rotating the sub-group elements allocation within the cell such that spectrally adjacent resonators are either two elements apart vertically and one horizontally (in a square lattice) or vice versa.
	
\begin{figure*}[t]
		\centering
		\includegraphics[]{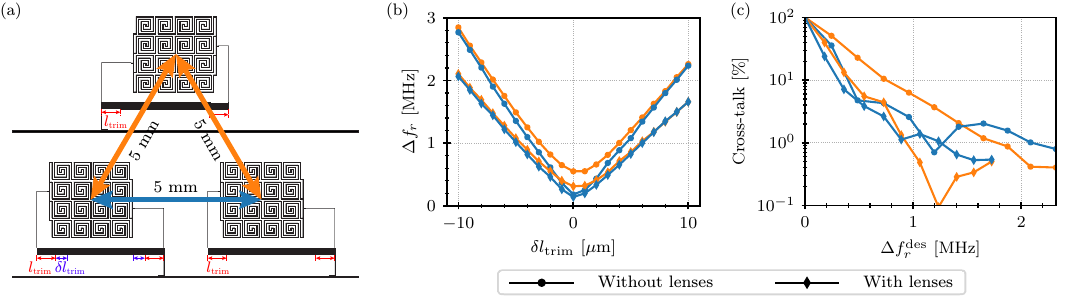}
		\caption{Cross-talk assessment between neighboring detectors. Sub-figure (a) shows the two possible means of cross-talk between \SI{5}{\milli\meter}-pitched hexagonally-packed spiral array resonators: laterally (blue) or transversely (orange). Sub-figure (b) shows the frequency offset of the two coupled resonators when compared to the isolated one for $\delta{}l_\mathrm{trim}=\SI{0}{\micro\meter}$. Sub-figure (c) quantifies the cross-talk between two neighboring resonators.}
		\label{fig:xtalk}
	\end{figure*}

	To quantify the level of cross-talk, we have simulated with Sonnet \cite{Sonnet} pairs of neighboring resonators where one is kept unaltered and the other is slightly de-tuned \cite{Noroozian2012,Yates2014}. As depicted in \cref{fig:xtalk}(a), the pairs of resonators analyzed are separated by a 5 mm pitch either laterally or transversely due to the hexagonal packing employed to maximize the array filling factor. Due to the electromagnetic interaction, the resonators become coupled and affect each other. To assess the cross-talk level, we have introduced a small change $\delta{}l_\mathrm{trim}$ affecting the length of the fingers of the interdigitated capacitor of one resonator. The frequency separation between the resonances of the coupled resonators is illustrated in \cref{fig:xtalk}(b). As expected, even when the resonators are exactly the same ($\delta{}l_\mathrm{trim}=0$), the two resonances are slightly different. The cross-talk level can be calculated from the ratio of resonance frequency change with respect to the nominal (isolated resonator) of the fixed resonator over the change for the varying one. The cross-talk level is plotted in \cref{fig:xtalk}(c) as a function of the design frequency spacing $\Delta{}f_r^\mathrm{des}$. The resonators cross-talk less than 10\% if they are spaced more than \SI{1}{\mega\hertz}, in both orientations for a 5 mm pitch. If the resonators are placed under the Si lens array, the cross-talk decreases to $\sim1\%$ for \SI{1}{\mega\hertz}. According to this, only neighboring resonators closer than \SI{1}{\mega\hertz} could incur about 10\% cross-talk, which by design are 5.5\% of them as shown in \cref{fig:MKIDshuffling}.

\begin{figure*}[t]
		\centering
		\includegraphics[]{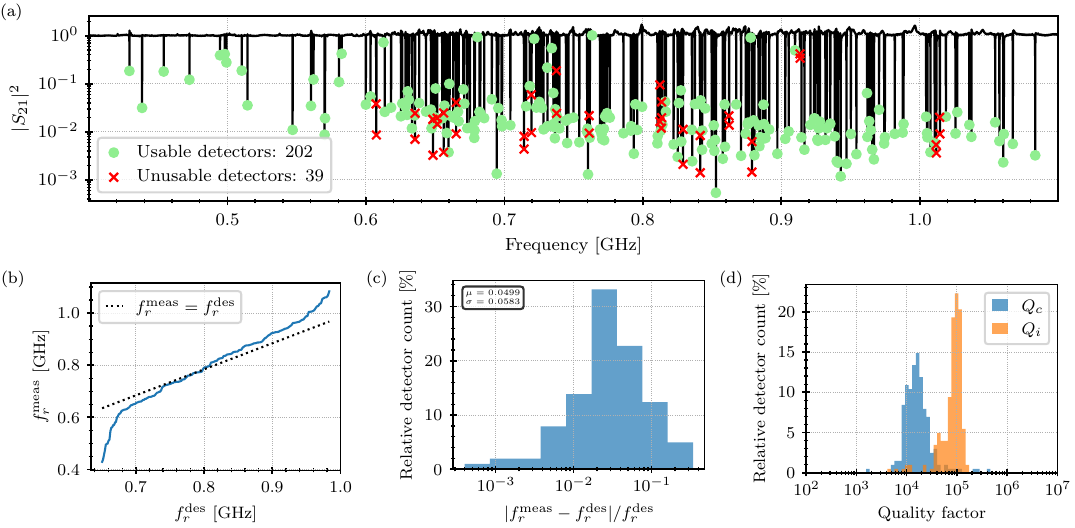}
		\caption{Panel (a) shows the transmission of the microwave readout of the 253 pixel wafer array measured at \SI{100}{\milli\kelvin} and normalized by the transmission measured at \SI{450}{\milli\kelvin}, where the resonators are too shallow due to the conduction loss. The total number of detected resonators is 241, giving a yield of $241/253\approx95\%$. Only the 202 resonators with neighboring resonances further than 20 half-power bandwidths are considered usable. Panel (b) illustrates the measured resonance frequencies as a function of the respective design frequencies. Panel (c) showcases the relative resonance frequency error with respect to the design. Panel (d) contains histograms of the coupling and internal quality factors.}
		\label{fig:wafer_measurements}
	\end{figure*}

	The large format demonstrator was measured at 100 mK bath temperature, but we limited the characterization to the microwave readout transmission. As depicted in \cref{fig:wafer_measurements}(a), we found 241 resonators out of 253, resulting in a 95\% detector yield; of which 39 are deemed unusable due to their close proximity (less than 20 half-power bandwidths) to a neighboring resonance. As shown in \cref{fig:wafer_measurements}(b) and (c), the measured resonance frequencies $f_r^\mathrm{meas}$ were found to deviate from the design values $f_r^\mathrm{des}$ an average of 5.0\% and with a standard deviation of 5.8\%. The trend of measured resonance frequencies with respect to the designed ones is reminiscent of the cross-talk analysis shown in \cite{Noroozian2012}. However, the seemingly large remnant cross-talk between resonators is likely caused by fabrication imperfections (film thickness variations, limited lithographical precision, etc.) and not \emph{ab initio} cross-talk as can be seen in \cref{fig:MKIDshuffling}, where the color of the different resonators indicate the designed spectral distance of immediately neighboring resonators at 5 mm pitch. Finally, the 202 usable resonators yield an average internal quality factor of $\overline{Q_i}\approx\num[mode=text]{1.6e5}$ and an average coupling quality factor of $\overline{Q_c}\approx\num[mode=text]{2.3e4}$ as indicated in \cref{fig:wafer_measurements}(d). 
	
	\section{Conclusions and Outlook}
	
	In this paper we have shown the design and measurements of spiral inductors Ti/Al lumped element resonators as dual-polarized absorber-coupled MKIDs for millimeter-wavelength radiation. The electromagnetic design simulations indicate that the lens-coupled spiral absorbers provide a lens aperture efficiency of more than 70\% for an octave bandwidth for the $4\times4$ spiral array absorber and for 10\% relative bandwidth for the single spiral absorber. First, a 9 detector chip is fabricated in order to test the two absorber designs. To ensure that only millimeter-wavelength radiation emanating from the calibrated blackbody is captured by the detectors, we designed and fabricated an 85 GHz Nb Fabry-Pérot band-pass filter that sits on one of the lids of a specifically designed chip holder with a \diameter3 cm aperture and otherwise light-tight. While the spiral array yields a better simulated optical response, we found that the single spiral absorber is more responsive than the spiral array absorber, likely due to the smaller active volume. Nonetheless, both detectors respond about \SI{1}{(\hertz\per\hertz)\per\micro\kelvin} and the PSD showcases an optical-loading-dependent quasi-particle roll-off with a Lorentzian shape, indicating photon-noise limited performance. We give a first estimate of the sensitivity with a NET of \SI{1}{\milli\kelvin\per\sqrt{\hertz}} at \SI{1}{\kilo\hertz} audio frequency, which is on par with other millimeter-wavelength cameras like NIKA \cite{NIKA1}. With these findings, we designed and fabricated a millimeter-wavelength $\diameter$4'' camera prototype featuring 253 dual-polarized spiral array MKIDs read out from a single line. Most of the detectors were found (95\% detector yield) and 80\% of the total detector count were deemed usable (spectrally distant), for which we performed statistics. We report an average internal quality factor of $\overline{Q_i}\approx\num[mode=text]{1.6e5}$, an average coupling quality factor of $\overline{Q_c}\approx\num[mode=text]{2.3e4}$, a resonance frequency scatter of 5.8\% and a mean frequency offset with respect to the design of 5.0\%. We attribute these frequency deviations to fabrication imperfections, such as thickness inhomogeneities in the superconducting film and the limited precision of optical lithography. 
	
	Follow-up work includes the integration of the detectors with the lenses, which are not expected to limit the sensitivity (e.g.\ \cite{Baselmans2022,Dabironezare2025}), especially if their backside is kept crystalline. We will carefully characterize the NEP and optical efficiency of these detectors including the lenses following the formalism for multi-moded detectors reported by Dabironezare \emph{et al.} \cite{Dabironezare2025}. To this end, we need to find out the cause of the high $1/f$ noise in our setup, and ameliorate it. Possible causes include electrical, thermal and mechanical time-dependent processes, as well as Two-Level Systems noise. Furthermore, we need to characterize our quasi-optical filters at \SI{4}{\kelvin} and include a box-in-a-box light-tight setup \cite{Baselmans2012} to absolutely quantify the blackbody power impinging on the detectors. Possible improvements in the sensitivity will come from using NbTiN for the capacitors to reduce the stray-light response and noise \cite{Barends2009,Janssen2013}. Lastly, an alternative route to absorber-coupled MKIDs is to employ antenna-coupled ones, where the quasi-optical coupling and detection mechanisms can be optimized largely independently from each other.

	\section*{Acknowledgment}
	
	The authors would like to thank J.~J.~A.~Baselmans, S.~O.~Dabironezare, J.~Martín-Pintado, M.~Calvo, and A.~Monfardini for the useful discussions and suggestions; and R.~Ferrándiz for his support in designing and fabricating the light-tight chip holder.
	
	\bibliographystyle{IEEEtran.bst}
	\bibliography{refs}
	
	\newpage
	
	\begin{IEEEbiography}[{\includegraphics[width=1in,height=1.25in,clip,keepaspectratio]{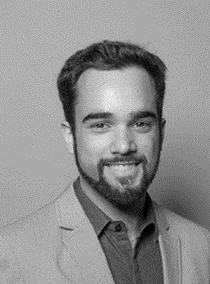}}]{Alejandro Pascual Laguna}
		was born in Madrid, Spain, in 1992. He received the B.Sc. degree in telecommunications engineering from ICAI School of Engineering, Universidad Pontificia Comillas, Madrid, Spain, in 2014 after spending an exchange year at Chalmers University of Technology, Gothenburg, Sweden. He then obtained the M.Sc. (cum laude) and Ph.D. degrees in electrical engineering, respectively in 2016 and 2022, from the Delft University of Technology, Delft, The Netherlands. 
		
		From 2016 to 2023, he was with SRON, the Space Research Organization of the Netherlands, Leiden, The Netherlands; first as a Ph.D. candidate and then as a Scientist. From 2023, he is with CAB, the Astrobiology Center (CSIC-INTA), Torrejón de Ardoz, Spain, where he is currently a Juan de la Cierva fellow. His research interests include on-chip solutions for efficient broadband (sub-)millimeter wavelength imaging spectrometers and polarimeters based on ultra-sensitive Kinetic Inductance Detectors.
	\end{IEEEbiography}
	
	\begin{IEEEbiography}[{\includegraphics[width=1in,height=1.25in,clip,keepaspectratio]{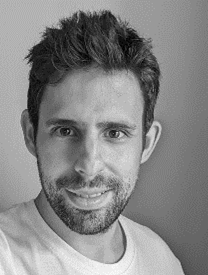}}]{Victor Rollano}
		was born in Madrid, Spain in 1988. He received his B.Sc. (2014) and M.Sc. (2015) degrees in physics from the Universidad Complutense de Madrid (Spain), and his Ph.D. degree from the same university and IMDEA Nanoscience in 2019. His doctoral research focused on the dynamics of superconducting vortices in engineered pinning landscapes created by magnetic and superconducting nanostructures.
		
		From 2020 to 2021, he was a postdoctoral researcher with the Quantum Materials and Devices Group at the Instituto de Nanociencia y Materiales de Aragón (CSIC), where he contributed to the development of quantum processing units based on molecular spins. He was awarded a Margarita Salas Fellowship in 2021 and a Marie Skłodowska-Curie Fellowship in 2022 to pursue research on hybrid quantum devices combining nitrogen-vacancy ensembles in diamond with superconducting qubits, carried out at the University of Science and Technology of China (USTC) within the Hybrid Quantum Devices Group. He is currently with the Centro de Astrobiología (CSIC-INTA), where his work focuses on the development of Kinetic Inductance Detectors for dark matter detection experiments.
	\end{IEEEbiography}
	
	\begin{IEEEbiography}[{\includegraphics[width=1in,height=1.25in,clip,keepaspectratio]{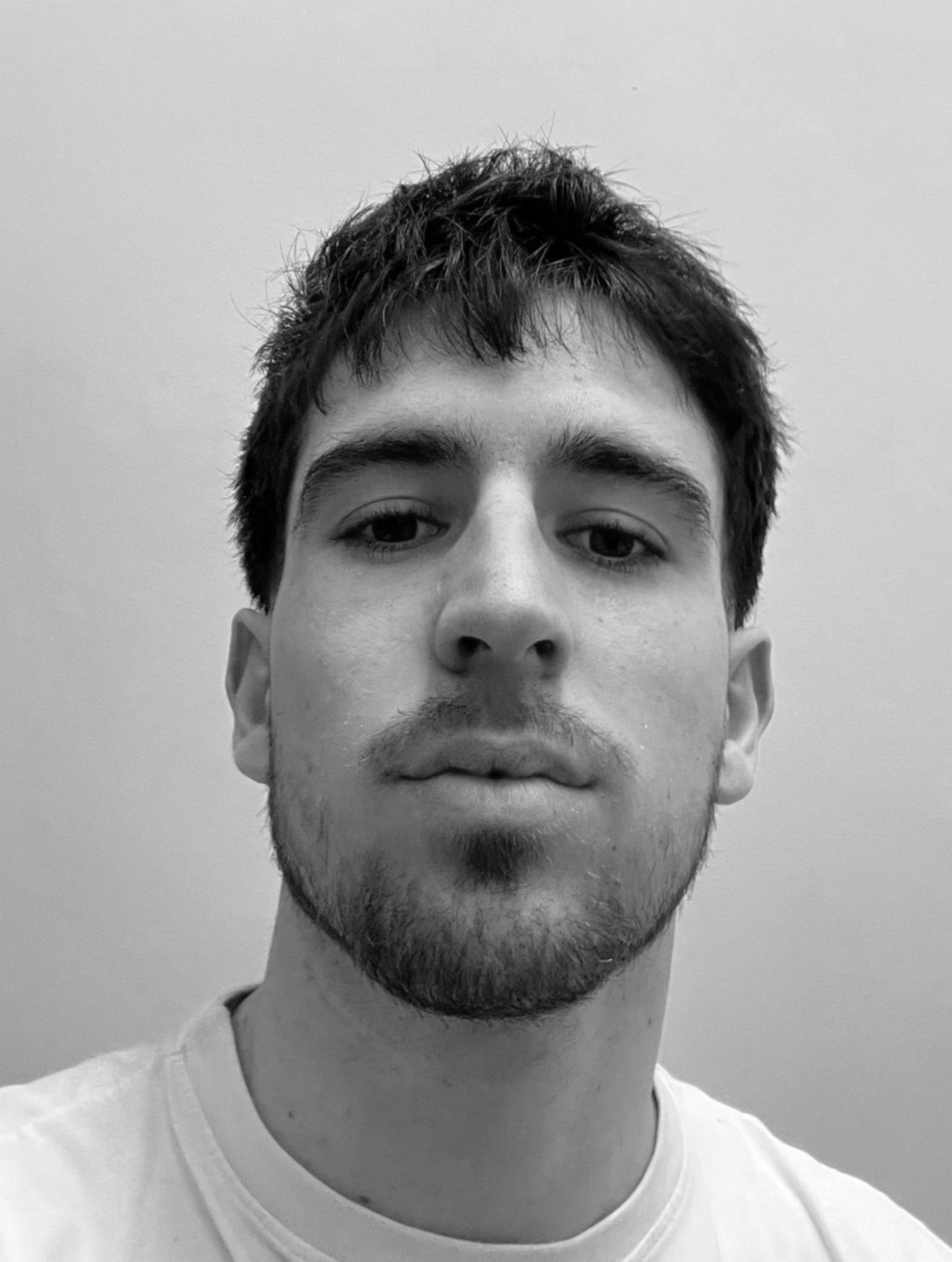}}]{Aimar Najarro-Fiandra}
		was born in Alcalá de Henares, Spain, in 2002. He received the B.Sc. degree in Physics and Space Instrumentation in 2025 from the University of Alcalá, Alcalá de Henares, Spain. He is currently pursuing the M.Sc. degree in Applied Physics at Delft University of Technology, Delft, The Netherlands.
		
		During his bachelor’s thesis, he was an intern at the Centro de Astrobiología (CSIC–INTA), Torrejón de Ardoz, Spain, where he was involved in research on Kinetic Inductance Detectors, cryogenics, and device fabrication. His research interests include superconducting devices and space instrumentation.
	\end{IEEEbiography}
	
	\begin{IEEEbiography}[{\includegraphics[width=1in,height=1.25in,clip,keepaspectratio]{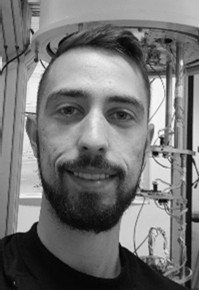}}]{David Rodriguez}
		was born in Zamora, Spain, in 1996. He received the B.Sc. degree in physics and M.Sc. degree in Condensed Matter Physics from the Universidad Autónoma de Madrid, Madrid in 2018 and 2019, respectively. He is currently pursuing the Ph.D. degree in the field of superconducting circuits for quantum technologies and quantum sensing with the Centro de Astrobiología (CSIC-INTA), Madrid under the supervision of Alicia Gomez.
		
		During his B.Sc. degree he joined the Department of Condensed Matter Physics at Universidad Autónoma de Madrid as a research assistant, where he was involved in quantum transport measurements. He later joined the Advanced Nanomaterials and Devices at Eindhoven University of Technology to conduct his master’s thesis. In 2021 he joined Alicia Gomez’s group as a research assistant and later became a Ph.D. student.
		
		His research focuses on the design of superconducting resonators for quantum and space applications. He is also involved in the development and control of qubits in collaboration with Instituto de Nanociencia y Materiales de Aragón (INMA) and Universidad de Zaragoza.
	\end{IEEEbiography}
	
	\begin{IEEEbiography}[{\includegraphics[width=1in,height=1.25in,clip,keepaspectratio]{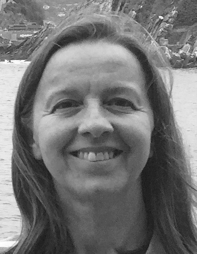}}]{Maria Teresa Magaz}
		was born in Madrid, Spain in 1965. She obtained her B.Sc. (1986) and M.Sc. (1988) degrees in Organic Chemistry from Universidad Autónoma de Madrid and the M.Sc. degree in Occupational Health and Safety in 2010.
		
		From 2010 to 2015, she was a Research Assistant Technician at the Institute of Technology La Marañosa (ITM) in Madrid, working on optical photolithography for the development of PbSe IR detectors. Since 2017 she has been working as a Research Assistant Technician in the Centro de Astrobiología (CSIC-INTA) in collaboration with IMDEA-Nanociencia in Madrid. Her research focuses on processing and optimization of nano- and micro-fabrication of ultrasensitive detectors for astronomy and quantum technologies. Her expertise includes sputtering and e-beam evaporation deposition in high vacuum conditions, dielectric deposition using Atomic Layer Deposition (ALD), optical and electrical lithography, wet and dry etching techniques and fabrication characterization.
	\end{IEEEbiography}
	
	\begin{IEEEbiography}[{\includegraphics[width=1in,height=1.25in,clip,keepaspectratio]{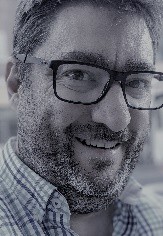}}]{Daniel Granados}
		was born in Madrid, Spain, in 1977. He received the B.Sc. (2001) and M.Sc. (2002) degrees in Physics from Universidad Autónoma de Madrid, Spain, and the Ph.D. degree in 2006 while working at the Spanish National Research Council (IMN-CNM-CSIC). In 2005 he was a Visiting Scientist with the Nano-Optics Group, Heriot-Watt University, Edinburgh, U.K. In 2006 he joined the Quantum Information Group, Toshiba Research Europe Ltd., Cambridge, U.K., where he worked on nanophotonic devices for quantum optics and cavity quantum electrodynamics and collaborated with the Semiconductor Physics Group, Cavendish Laboratory, University of Cambridge.
		
		He joined IMDEA Nanoscience, Madrid, Spain, in 2009. He is currently Executive Director of Scientific Infrastructures and Head of the Quantum NanoDevices Group. He led the creation of the NanoFabLab cleanroom facility, operational since 2014, supporting advanced micro- and nanofabrication and semiconductor device research. His research interests include nanofabrication, superconducting detectors, devices based on two-dimensional materials, photonics, nano-optics, and quantum technologies.
		
		Dr. Granados has authored 95 peer-reviewed journal publications, contributed to 80 international conference presentations, and holds two patents. He has been Principal Investigator in 18 competitive research projects or contracts and has contributed to 25 national and international research projects, including European programs. He served as Vice-President of the Electronic Materials Division of the International Union for Vacuum Science, Technique and Applications (IUVSTA). He is a member of the Executive Committee of the Fundación Círculo de Tecnologías para la Defensa y la Seguridad and Director of the Cluster de Innovación Tecnológica y Talento en Semiconductores (CITT Semiconductors) of the Madrid Region. He represents the Comunidad de Madrid in the European Semiconductor Regions Alliance (ESRA) and participates in advisory and working groups at regional, national, and European levels.
	\end{IEEEbiography}
	
	\begin{IEEEbiography}[{\includegraphics[width=1in,height=1.25in,clip,keepaspectratio]{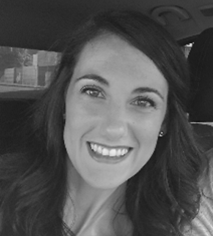}}]{Alicia Gomez}
		was born in Madrid, Spain in 1986. She obtained her B.Sc. (2009) and M.Sc. (2010) in Physics from the Universidad Complutense de Madrid (Spain). She obtained her Ph.D. from the Universidad Complutense of Madrid in 2013, focused on superconducting vortex dynamics.
		
		She was a visiting researcher at the University California-Davis (USA) in 2010 and at the University of Antwerpen (Belgium) in 2012. Since 2013, she is a postdoctoral researcher at Centro de Astrobiología (CSIC-INTA) in Spain, where she is in charge of the development of superconducting Kinetic Inductance Detectors for space and ground based instrumentation. She is member of international collaborations such as the NIKA2 camera, installed at the IRAM 30 m telescope in Granada, and the KISS camera, for the Canary Island telescope. She was also part of the CORE proposal submitted to the call for the M5 mission of the ESA Cosmic Vision. She has authored or co-authored more than 40 peer-reviewed papers. Her experience spans from the design and fabrication of superconducting electronics in cleanroom to the electronic characterization of superconducting devices at cryogenic temperatures with and without magnetic fields. 
		
		Dr. Gomez was awarded with a Spanish MICINN Juan de la Cierva-Incorporación postdoctoral fellowship in 2018 to lead the development of Kinetic Inductance Detectors at Centro de Astrobiología (CSIC-INTA), Spain for two years.
	\end{IEEEbiography}

\end{document}